\documentclass[aps,twocolumn,superscriptaddress,showpacs]{revtex4}

\usepackage[dvipdf]{graphicx}

\usepackage{amssymb}

\usepackage{amsbsy}

\usepackage{latexsym}

\begin{document}

\title{Transport control in deterministic ratchet system}

\author{Woo-Sik \surname{Son}}

\email{dawnmail@sogang.ac.kr} 

\affiliation{Department of Physics, Sogang University, Seoul 121-742, Korea}

\affiliation{National Creative Research Initiative Center for Controlling Optical Chaos,
  Pai-Chai University, Daejeon 302-735, Korea}

\author{Jung-Wan \surname{Ryu}} \affiliation{Department of Physics, Pusan National University, Busan 609-735, Korea}

\author{Dong-Uk \surname{Hwang}} \affiliation{Department of Biomedical Engineering, University of Florida, Florida 32611-6131}

\author{Soo-Young \surname{Lee}} \affiliation{National Creative Research Initiative Center for Controlling Optical Chaos,
  Pai-Chai University, Daejeon 302-735, Korea}

\author{Young-Jai \surname{Park}} 

\email{yjpark@sogang.ac.kr}

\affiliation{Department of Physics, Sogang University, Seoul 121-742, Korea}

\author{Chil-Min \surname{Kim}} 

\email{chmkim@mail.pcu.ac.kr}

\affiliation{National Creative Research Initiative Center for Controlling Optical Chaos,
  Pai-Chai University, Daejeon 302-735, Korea}

\date{\today}

\begin{abstract}

We study the control of transport properties in a deterministic inertia ratchet system
via the extended delay feedback method.
A chaotic current of a deterministic inertia ratchet system is controlled to a regular current by stabilizing 
unstable periodic orbits embedded in a chaotic attractor of the unperturbed system. 
By selecting an unstable periodic orbit, which has a desired transport property, 
and stabilizing it via the extended delay feedback method,
we can control transport properties of the deterministic inertia ratchet system.
Also, we show that the extended delay feedback method can be utilized 
for separation of particles
in the deterministic inertia ratchet system as a particle's initial condition varies.

\end{abstract}

\pacs{05.40.-a, 05.45.Gg, 05.45.Pq, 05.60.Cd}

\maketitle

\section{INTRODUCTION}

The ratchet effect, i.e., a directional motion of a particle using unbiased fluctuations,
has attracted much attention in recent years \cite{Reimann,Astumian}.
An early motivation in this field is to explain an underlying mechanism of molecular motors
which transport molecules in the absence of appropriate potential and thermal gradients \cite{Astumian2}.
Lately, the ratchet effect has been studied theoretically and experimentally
in many different fields of science, e.g., asymmetric superconducting quantum interference devices \cite{Zapata},
quantum Brownian motion \cite{Grifoni}, Josephson-junction arrays \cite{Lee}, 
application for separation of particles \cite{Rousselet}, etc.
It has been known that two conditions should be met to obtain the ratchet effect \cite{Reimann}.
First, a system has to be in a non-equilibrium state by a correlated stochastic \cite{Bartussek}
or a deterministic perturbation \cite{Magnasco}.
Second, the breaking of the spatial inversion symmetry is required.
In doing so, an asymmetric periodic potential, named the ``ratchet potential'', is introduced.

Recently, several works concerning the control of ratchet dynamics have been presented.
The applying of a weak subharmonic driving in a deterministic inertia ratchet system
was used to enlarge the parameter ranges where regular currents are observed \cite{Barbi2},
and the signal mixing of two driving forces was considered to control transport properties
in a overdamped ratchet system \cite{Savelev}.
Also, the effect of time-delayed feedback on the ratchet system has been studied.
The anticipated synchronization was observed in delay coupled inertia ratchet systems \cite{Mateos2}
and the stabilization of chaotic current to low-period orbits was presented,
using time-delayed feedback methods, 
in the deterministic inertia ratchet system \cite{Son2}.

Starting with the work of Ott, Grebogi, and Yorke \cite{Ott},
various methods for controlling chaotic dynamics have been developed \cite{Shinbrot}.
Particularly, Pyragas proposed a simple and efficient method,
which utilizes a control signal with a difference between the present state of the system and
the previous state delayed by the period of an unstable periodic orbit (UPO) \cite{Pyragas}.
This method is noninvasive in the sense that the control signal vanishes when the targeted UPO embedded in
a chaotic attractor is stabilized.
Some limitations on the Pyragas method have been reported \cite{Ushio} and 
the modifications of the Pyragas method have been proposed to improve its efficiency \cite{Socolar,Kittel}.
Socolar {\it et al}. presented a method utilizing information from
many previous states of the system, and this method is called
the {\it extended time-delay autosynchronization} or the {\it extended delay feedback} (EDF) \cite{Socolar}.
The stability and analytical properties of a delayed feedback system
have been investigated \cite{Bleich}.

In this paper, we aim to control transport properties of the deterministic inertia ratchet system.
For this purpose, we have controlled a chaotic current of the system to a regular current by stabilizing an unstable periodic orbit
which has a desired mean velocity, via the EDF method. 
Also, we have shown that the EDF method can be utilized 
for separation of particles
in the deterministic inertia ratchet system as a particle's initial condition varies.
The rest of the paper is organized as follows.
In Sec. II, we have shown transport properties of the unperturbed deterministic inertia ratchet system.
In Sec. III, the system controlled via the EDF method has been presented 
and the linear stability analysis of a periodic orbit in the presence of the EDF has been considered. 
In Sec. IV, via the EDF method, we have shown achievements of the desired transport properties of the system and a
separation of particles has also been presented as varying particle's initial condition.
The paper finishes with conclusions in Sec. V.

\section{DETERMINISTIC RATCHET SYSTEM}

The deterministic inertia ratchet system is
written as the following dimensionless equation,

\begin{equation}\label{eq:1}
\ddot{x}+b\dot{x}+V^{\prime}(x)=a\cos(\omega_{0}t).
\end{equation}

\begin{figure}[b!]
\includegraphics{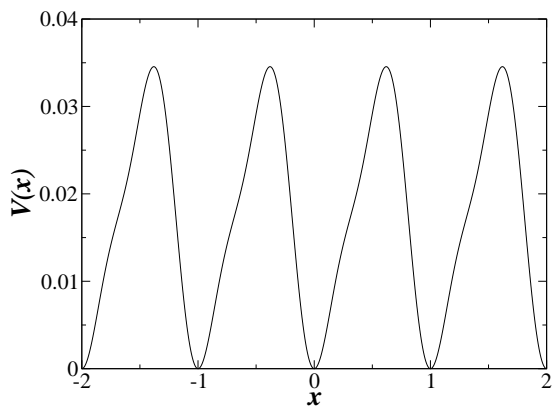}
\caption{Asymmetric periodic potential, i.e., the ratchet potential 
$V(x)=C-(\sin(2\pi(x-d))+\frac{1}{4}\sin(4\pi(x-d)))/4\pi^{2}\delta$
with $d=-0.19$, $\delta = \sin(2{\pi}|d|) + \sin(4{\pi}|d|)$, and
$C=-(\sin(2\pi{d})+0.25\sin(4\pi{d}))/4\pi^{2}\delta$.}
\label{fig:fig1.eps}
\end{figure}

\noindent Here, $V^{\prime}(x)$ denotes the derivative with respect to $x$ and $b$ is the friction coefficient.
$\omega_{0}$ and $a$ are the frequency and amplitude of the driving force, respectively.
Figure \ref{fig:fig1.eps} shows an asymmetric periodic potential,
i.e., the ratchet potential $V(x)$ described by

\begin{equation}\label{eq:2}
V(x)=C-\Big(\sin\big(2\pi(x-d)\big)+\frac{1}{4}\sin\big(4\pi(x-d)\big)\Big)/4\pi^{2}\delta,
\end{equation}

\noindent where $d$, $\delta$ and $C$ are introduced so that the
ratchet potential has a minimum at $x=0$ with $V(0)=0$. 

This system exhibits both regular and chaotic behaviors, depending
on parameters $(a,b,\omega_{0})$ \cite{Mateos,Barbi,Son}.
In this paper, we vary only the parameter $a$, and set $b=0.1$ and $\omega_{0}=0.67$.
General transport properties of the deterministic inertia ratchet system 
are shown in Fig. \ref{fig:fig2.eps}.
In Fig. \ref{fig:fig2.eps}(a), we plot a bifurcation diagram of
the stroboscopic recording of a particle's velocity at $t=kT$,
where $k$ is a positive integer and $T$ is the period of
the driving force. 
The mean velocity of the system as a function of the parameter $a$
is depicted in Fig. \ref{fig:fig2.eps}(b).
As shown in Fig. \ref{fig:fig2.eps}(b), 
multiple current reversals occur as the amplitude of the driving force is varied.
It has also been observed that the current reversal is related to
a bifurcation from chaotic to regular regime \cite{Mateos}
and that the type-I intermittency exists in this bifurcation \cite{Son}. 

When the system exhibits a regular behavior, 
the time required for a particle to move from one well of the potential to another 
is commensurable with the period of the driving force.
Hence, the mean velocity of a regular current shows a locking phenomenon given as follows:

\begin{figure}[t!]
\includegraphics{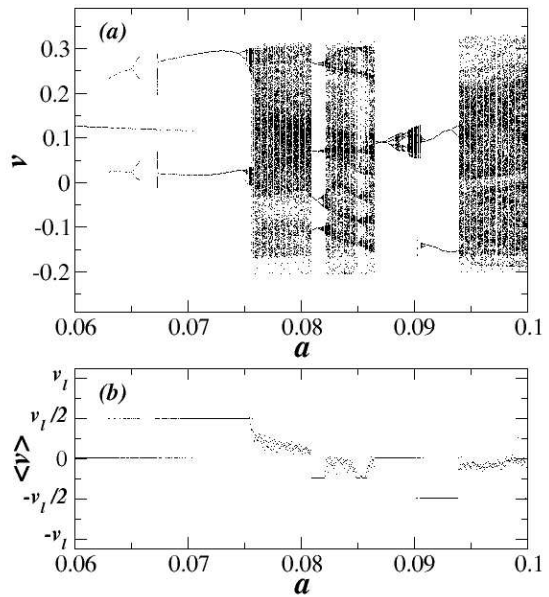}
\caption{Bifurcation diagrams as a function of $a$ at $b=0.1$ and $\omega_0 = 0.67$. 
In the region from $a=0.063$ to $0.071$, coexisting attractors are found; 
(a) the stroboscopic recording of particle velocity,
(b) the mean velocity of current.}
\label{fig:fig2.eps}
\end{figure}

\begin{equation}\label{eq:3}
\langle v \rangle=\frac{n}{m}\frac{L}{T}=\frac{n}{m}\frac{\omega_{0}}{2\pi}L=\frac{n}{m}v_{l},
\end{equation}

\noindent where $L$ is the spatial period of the ratchet potential 
($L=1$, as shown in Fig. \ref{fig:fig1.eps}, then $v_{l}=\frac{\omega_{0}}{2\pi}$),
$T$ is the time period of the driving force,  
and $\frac{n}{m}$ is an irreducible fraction $(n,m \in Z)$ \cite{Barbi}.
$v_{l}$ is the fundamental locking velocity corresponding to 
a particle's current which advances one well of the ratchet potential in a positive direction with the period of the driving force.
As shown in Fig. \ref{fig:fig2.eps}, the system exhibits regular behaviors in some parameter ranges;
a period-1 orbit with $\langle{v}\rangle = 0$ ($a=0.06$),
a period-2 orbit with $\langle{v}\rangle = \frac{1}{2}v_{l}$ ($a=0.074$), 
a period-4 orbit with $\langle{v}\rangle = -\frac{1}{4}v_{l}$ ($a=0.081$), and
a period-2 orbit with $\langle{v}\rangle = -\frac{1}{2}v_{l}$ ($a=0.092$).
When the system shows a chaotic behavior, the mean velocity of the chaotic current is almost zero averaged.

\begin{figure}[t!]
\includegraphics{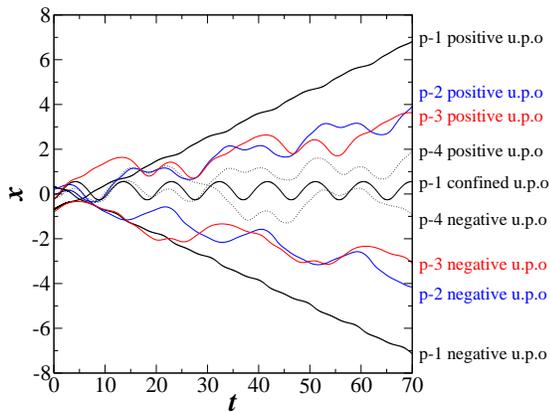}
\caption{(Color online) Unstable periodic orbits.
period-1 (black line), period-2 (blue line), period-3 (red line), and period-4 orbits (dotted black line)
obtained from the unperturbed system at $a=0.083$.}
\label{fig:fig3.eps}
\end{figure}

It is worthy of note that there are various UPOs embedded in
a chaotic attractor of the unperturbed system and that their mean velocities agree with Eq. (\ref{eq:3}).
By stabilizing an UPO that has a desired mean velocity (i.e., written by specific $n$ and $m$ in Eq. (\ref{eq:3})),
we can achieve a desired regular transport of the deterministic inertia ratchet system 
instead of the zero averaged chaotic current.
In this system, the periodic orbit is defined by
$\big(\tilde{x}(t),v(t)\big)=\big(\tilde{x}(t+\tau),v(t+\tau)\big)$,
where $\tilde{x}(t)=x(t)(mod$ $1)$ and $\tau$ is the time period of the orbit. 
We are interested in stabilizing some targeted UPOs among various UPOs that agree with Eq. (\ref{eq:3}).
In Fig. \ref{fig:fig3.eps}, we have shown several UPOs, which are desired to be stabilized, 
located by the Newton method.
For period-$n$ orbits, we consider two UPOs that have the same period $\tau=nT=2n\pi/\omega_0$
with different mean velocities,
where $n=1,2,3,4$:
one is a positive current with the mean velocity ${\langle}v{\rangle}=\frac{1}{n}v_{l}$ and 
the other is a negative current with the mean velocity ${\langle}v{\rangle}=-\frac{1}{n}v_{l}$.
Particularly, period-1 orbits include an oscillating orbit confined in one well of the potential with the mean velocity 
${\langle}v{\rangle}=0$.
By selecting a specific UPO and stabilizing it via the EDF method, 
we can easily control transport properties of the deterministic inertia ratchet system.

\section{LINEAR STABILITY ANALYSIS OF PERIODIC ORBITS}

The deterministic inertia ratchet system controlled by the EDF method is
described as

\begin{equation}\label{eq:4}
\ddot{x}+b\dot{x}+V'(x)=a\cos(\omega_{0}t)+F,
\end{equation}

\noindent where $F$ is a control signal, i.e., the delayed feedback described by the particle's present velocity
and the previous velocities delayed by multiples of the period of UPO.
$F$ is denoted by

\begin{equation}\label{eq:5}
F= K\big((1-R)\sum_{m=1}^{\infty}R^{m-1}\dot{x}(t-m\tau)-\dot{x}(t)\big),
\end{equation}

\noindent where $K$ is a strength of feedback, $\tau$ is a delay
time, which coincides with the period of the targeted UPO,
and $R$ $(0 \leq R < 1)$ is a parameter that adjusts the distribution of each
term's magnitude in the control signal.
When $R=0$, the EDF method
is the same as the Pyragas method, i.e.,
$F=K\big(\dot{x}(t-\tau)-\dot{x}(t)\big)$.

Now, let us consider the linear stability analysis of a periodic orbit in the presence of the EDF.
Let the small deviation from the periodic orbit $\boldsymbol{\xi_{0}}(t)$ be
$\delta\boldsymbol{\xi}(t)=\boldsymbol{\xi}(t)-\boldsymbol{\xi_{0}}(t)$.
According to the Floquet theory, $\delta\boldsymbol{\xi}(t)$ can be described as

\begin{equation}\label{eq:6}
\delta\boldsymbol{\xi}(t)=\sum_{k=1}^{N}C^{(k)}e^{(\lambda_{k}+i\omega_{k})t}\boldsymbol{u}_{k}(t),
\end{equation}

\noindent where $\lambda_{k}+i\omega_{k}$ is the Floquet exponent and 
$\boldsymbol{u}_{k}(t)=\boldsymbol{u}_{k}(t+\tau)$ is an eigenvector. 
$C^{(k)}$ is a constant and $N$ is the dimension of the Poincar\'{e} surface.
For one such mode, one can obtain the following deviation relation (dropping the index $k$).
After the period $\tau$ of the periodic orbit has passed,
the deviation is described as

\begin{eqnarray}\label{eq:7}
\delta\boldsymbol{\xi}(t+\tau) &=& \exp\big((\lambda+i\omega)(t+\tau)\big)\boldsymbol{u}(t+\tau)\\
                  &=& \exp\big((\lambda+i\omega)\tau\big)\delta\boldsymbol{\xi}(t)\equiv(\Lambda+i\Omega)\delta\boldsymbol{\xi}(t)\nonumber,
\end{eqnarray}

\noindent where $\Lambda+i\Omega$ is the Floquet multiplier.
When the delay terms are included, the phase space of the system becomes infinite dimensional
and the system has an infinite number of Floquet multipliers.
If the largest Floquet multiplier satisfies $|\Lambda_1+i\Omega_1| < 1$, i.e.,
the leading Floquet exponent $\lambda_1$ $(\lambda_{1} = \ln |\Lambda_1+i\Omega_1|/\tau)$ is less than zero,
thereby the targeted UPO is stabilized.

The time evolution of $\delta\boldsymbol{\xi}(t)$ is given by

{\setlength\arraycolsep{1pt}
\begin{eqnarray}\label{eq:8}
\delta\dot{\boldsymbol{\xi}} &=& \left(\begin{array}{cc} 0 & 1\\
-V''(x) & -b\end{array}\right)\delta\boldsymbol{\xi}(t) {} \\
&& {}+\left(\begin{array}{cc}
0 & 0\\
0 & 1\end{array}\right) K\big((1-R)\sum_{m=1}^{\infty}R^{m-1}\delta\boldsymbol{\xi}(t-m\tau)-\delta\boldsymbol{\xi}(t)\big), \nonumber
\end{eqnarray}}

\noindent where the matrix in the first term in Eq. (\ref{eq:8}) is the Jacobian of the
unperturbed system and the second term comes from the presence of the EDF.
The delayed terms in Eq. (\ref{eq:8}) can be eliminated
and consequently the time evolution of the small deviation from the periodic orbit
could be governed by

\begin{equation}\label{eq:9}
\delta\dot{\boldsymbol{\xi}}=\left(\begin{array}{cc}
0 & 1\\
-V''(x) &
-b-K\frac{1-\frac{1}{\Lambda+i\Omega}}{1-\frac{R}{\Lambda+i\Omega}}\end{array}\right)\delta\boldsymbol{\xi}(t)
=A\delta\boldsymbol{\xi}(t),
\end{equation}

\noindent where the ratio of geometric series has a constraint $\frac{R}{|\Lambda +i\Omega|}<1$
for convergence.
For an elimination of the delay terms, we use the relation

\begin{eqnarray}\label{eq:10}
\delta\boldsymbol{\xi}(t-n\tau)=(\Lambda+i\Omega)^{-n}\delta\boldsymbol{\xi}(t), & (n=1,2,3,\cdots). &
\end{eqnarray}
Eq. (\ref{eq:9}) requires information of the targeted UPO.
Hence, the Floquet multiplier is related to the eigenvalue problem of
the monodromy matrix $\Phi_{\tau}$, which satisfies

\begin{eqnarray}\label{eq:11}
\dot\Phi_{t}=A\Phi_{t}, & \Phi_{0}=I. &
\end{eqnarray}

\begin{figure}[t!]
\includegraphics{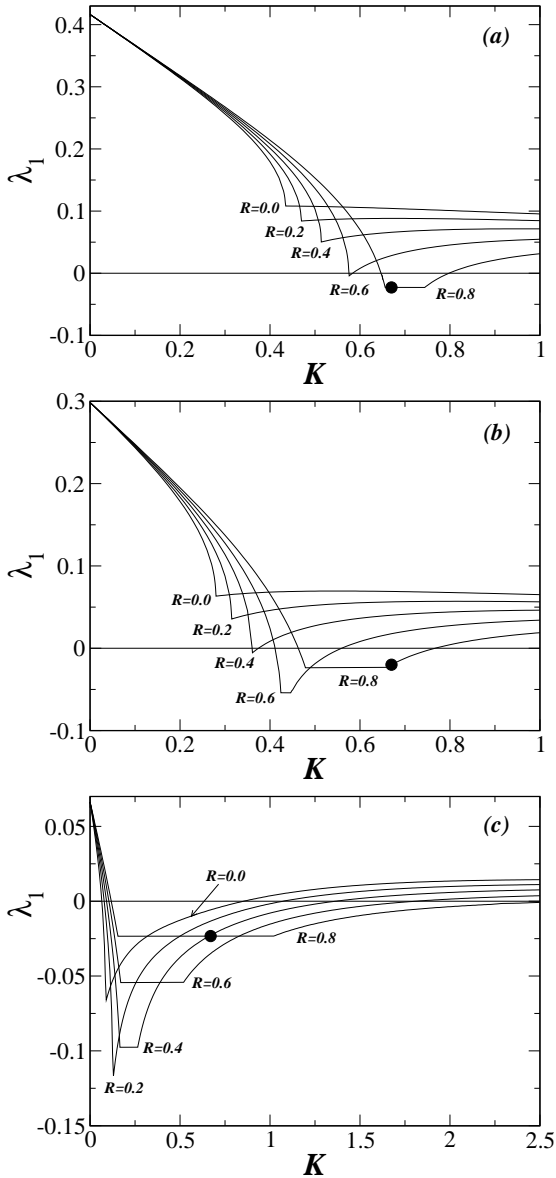}
\caption{The leading Floquet exponents of period-1 orbits;
(a), (b), and (c) exhibit the leading Floquet exponents for the positive, the negative,
and the confined currents as a function of $K$ for the given $R$, respectively.} \label{fig:fig4.eps}
\end{figure}

\noindent The eigenvalue of $\Phi_{\tau}$ defines the Floquet multiplier as
follows:

\begin{equation}\label{eq:12}
\det[\Phi_{\tau}-(\Lambda+i\Omega)I]=0.
\end{equation}

\begin{figure}[t!]
\includegraphics{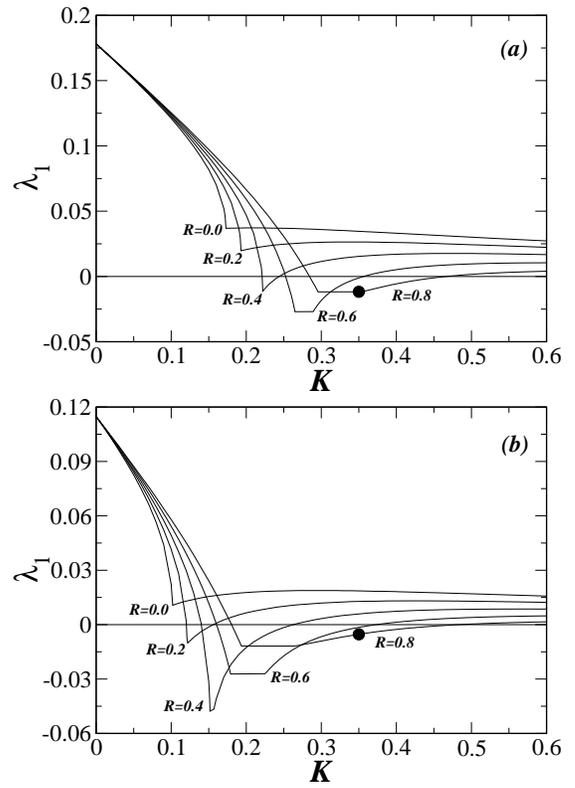}
\caption{The leading Floquet exponents of period-2 orbits;
(a) and (b) exhibit the leading Floquet exponents for the positive and the negative currents
as a function of $K$ for the given $R$, respectively.} \label{fig:fig5.eps}
\end{figure}

\begin{figure}
\includegraphics{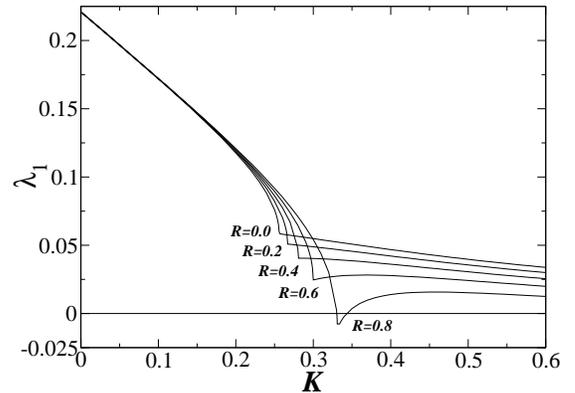}
\caption{The leading Floquet exponent of the period-3 positive orbit as a function of $K$ for the given $R$.} \label{fig:fig6.eps}
\end{figure}

\begin{figure}[t!]
\includegraphics{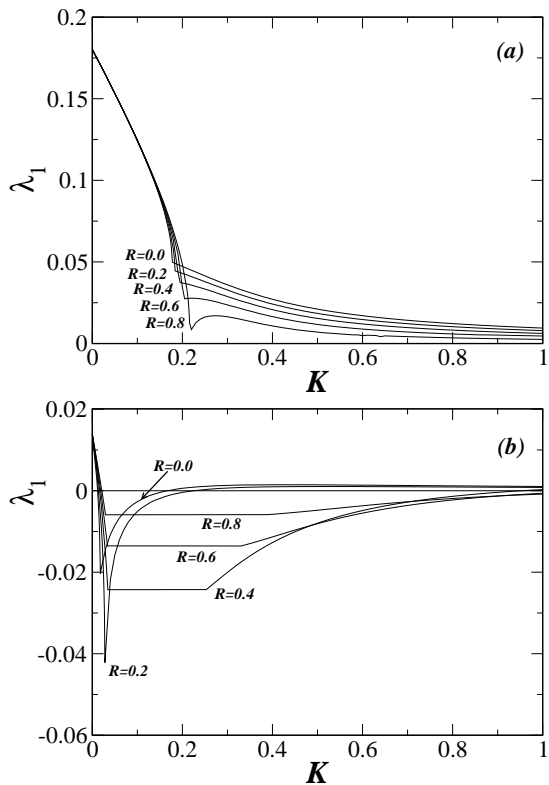}
\caption{The leading Floquet exponents of period-4 orbits;
(a) and (b) exhibit the leading Floquet exponents for the positive and the negative currents
as a function of $K$ for the given $R$, respectively.} \label{fig:fig7.eps}
\end{figure}

The Floquet multipliers are obtained by numerically
solving Eqs. (\ref{eq:9}),(\ref{eq:11}), and (\ref{eq:12}).
The results of leading Floquet exponents are shown in Figs. \ref{fig:fig4.eps}, \ref{fig:fig5.eps},
\ref{fig:fig6.eps}, and \ref{fig:fig7.eps} for period-1 ($\tau=T=2\pi/\omega_0$), period-2 ($\tau=2T$), 
period-3 ($\tau=3T$), and period-4 orbits ($\tau=4T$), respectively.
Each figure shows the leading Floquet exponent $\lambda_1$ 
as a function of the strength of feedback $K$ for different values of
the control parameter $R$.
The results tell us the stabilized region of control parameters $(K,R,\tau)$,
in which the targeted UPO is stabilized ($\lambda_1 < 0$).   

As shown in Fig. \ref{fig:fig4.eps}, 
the period-1 positive and the negative currents cannot be stabilized by the Pyragas method ($R=0$).
The positive current is stabilized at $R \geq 0.6$,
the negative current is stabilized at $R \geq 0.4$,
and the confined current, which only exists for the case of the period-1 orbits, can be stabilized even at $R=0$.
Such a phenomenon is roughly explained by the different degrees of instability 
of UPOs in the unperturbed system (see $\lambda_{1}$ values at $K=0$ in Fig. \ref{fig:fig4.eps}).
With the larger degree of instability in the unperturbed system, 
the UPO can be stabilized with a larger $R$.
The stabilized region of $K$ for the given $R$ increases as $R$ increases 
and if $R$ is not very large, then the minimum of $\lambda_1$ is deeper as $R$ increases.
The EDF method ($0<R<1$) is more effective than the Pyragas method ($R=0$).

Another interesting property of the EDF method is
a limitation on the minimum of $\lambda_1$ for given control parameters $R$ and $\tau$.
As shown in Fig. \ref{fig:fig4.eps}, 
the leading Floquet exponent is always larger than $\lambda^*$, where $\lambda^* = \ln(R)/\tau$.
$\lambda^{*}$ is given by the constraint on the ratio of geometric series.
In Eqs. (\ref{eq:8})-(\ref{eq:10}), the system gains the limit of geometric summation 
of the delayed feedback terms, which do not diverge, 
when the Floquet exponent is infinitesimally larger than $\lambda^*$.
$\lambda^*$ of the period-1 orbits ($\tau=T=2\pi/\omega_0$) is numerated as 
$\lambda^* \simeq -0.1716$ ($R=0.2$), $-0.0977$ ($R=0.4$),
$-0.0544$ ($R=0.6$), and $-0.0237$ ($R=0.8$).
If the result of the Floquet exponent obtained by solving Eqs. (\ref{eq:9}), (\ref{eq:11}), and (\ref{eq:12})
is far from $\lambda^*$ for given $R$ and $\tau$, then the shape of $\lambda_1(K)$ makes a narrow valley.
If the result is very close to $\lambda^*$,
then the shape of $\lambda_1(K)$ makes a valley with a flat bottom. 
In this case, the minimun of $\lambda_1$ is infinitesimally larger than $\lambda^*$ and
$\lambda_1$ is very slowly increasing as $K$ increases. 

\begin{figure}[t!]
\includegraphics{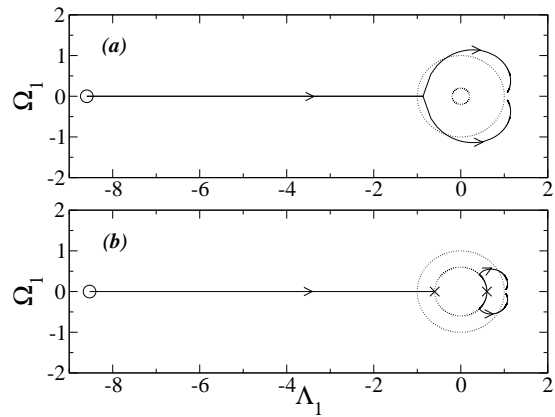}
\caption{The loci of the largest Floquet multiplier of period-2 negative orbit as $K$ varies from $0$ to $\infty$;
(a) loci at $R=0.2$ and (b) loci at $R=0.6$. The outer (inner) circle has a radius of $1$ ($R$).
The open circle and the crosses denote the location of the largest Floquet multiplier at $K=0$ and
the discontinuity, respectively. For $K \rightarrow \infty$, the largest Floquet multiplier approaches 
$(\Lambda_{1},\Omega_{1})=(1,0)$ in both cases of (a) and (b).}
\label{fig:fig8.eps}
\end{figure}

The leading Floquet exponents for period-2, period-3, and period-4 orbits 
are shown in Figs. \ref{fig:fig5.eps}, \ref{fig:fig6.eps}, and \ref{fig:fig7.eps}, respectively.
The results are qualitatively equivalent to those of period-1 orbits shown in Fig. \ref{fig:fig4.eps}. 
The period-2 positive current is stabilized at $R \geq 0.4$, and
the period-2 negative current is stabilized at $R \geq 0.2$.
$\lambda^*$ of the period-2 orbits ($\tau=2T=4\pi/\omega_0$) is numerated as 
$\lambda^* \simeq -0.0858$ ($R=0.2$), $-0.0488$ ($R=0.4$),
$-0.0272$ ($R=0.6$), and $-0.0118$ ($R=0.8$).
The period-3 negative current is stabilized at $R \geq 0.8$.
$\lambda^*$ of the period-3 orbit ($\tau=3T=6\pi/\omega_0$) is numerated as 
$\lambda^* \simeq -0.0572$ ($R=0.2$), $-0.0325$ ($R=0.4$),
$-0.0181$ ($R=0.6$), and $-0.0079$ ($R=0.8$).
The period-3 positive current cannot be stabilized by the EDF method
because the largest Floquet multiplier is a positive real value when the system is unperturbed.
It has been known that the delayed feedback method, including the EDF method, can stabilize
only a certain class of periodic orbits with a finite torsion \cite{Ushio}.
The period-4 positive current cannot be stabilized at $R \leq 0.8$ 
because it has a greater degree of instability $(\Lambda_{1} \simeq -862, \Omega_{1}=0)$ when $K=0$.
The period-4 negative current can be stabilized even at $R = 0$.
$\lambda^*$ of the period-4 orbits ($\tau=4T=8\pi/\omega_0$) is numerated as 
$\lambda^* \simeq -0.0429$ ($R=0.2$), $-0.0244$ ($R=0.4$),
$-0.0136$ ($R=0.6$), and $-0.0059$ ($R=0.8$).

\begin{figure}[t!]
\includegraphics{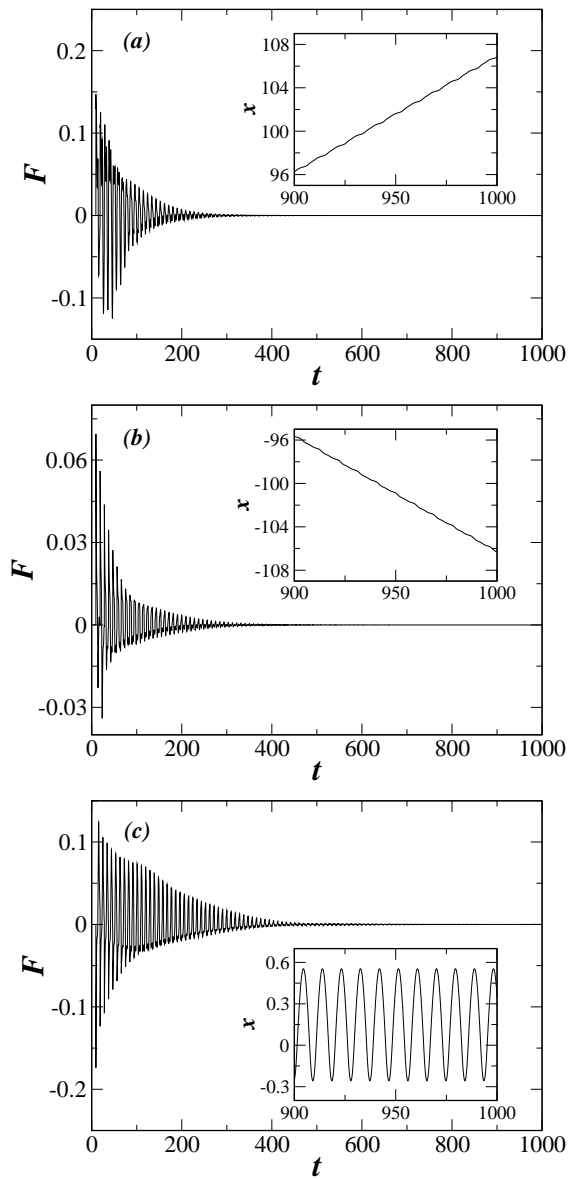}
\caption{Stabilized period-1 orbits (insets) and the dynamics of the control signal $F$  
at $K=0.67$, $R=0.8$, and $\tau=T=2\pi/\omega_0$;
(a) the positive current from the initial condition $(x_0,v_0)=(-0.35,0.2)$,
(b) the negative current from $(0.2,0.0)$,
and (c) the confined current from $(0.0,0.0)$.}
\label{fig:fig9.eps}
\end{figure}

All periodic orbits in Figs. \ref{fig:fig4.eps}, \ref{fig:fig5.eps}, \ref{fig:fig6.eps}, and \ref{fig:fig7.eps}
have common properties.
When $K=0$, their largest Floquet multipliers are real and negative so that 
the corresponding leading Floquet exponents satisfy $\omega_{1}={\pi}/{\tau}$.
It means that all UPOs flip their neighborhood within the period $\tau$ in the unperturbed system.
The variation of each orbit's largest Floquet multiplier depending on the feedback strength $K$ 
has a common aspect as shown in Fig. \ref{fig:fig8.eps}.
We plot the loci of the largest Floquet multiplier of period-2 negative orbit at $R=0.2$ (Fig. \ref{fig:fig8.eps}(a)),
and at $R=0.6$ (Fig. \ref{fig:fig8.eps}(b)).
Let us see Fig. \ref{fig:fig8.eps}(a).
As $K$ increases, the largest Floquet multiplier moves toward the zero point in remaining a negative
real value and crosses the unit circle so that the period-2 negative orbit is stabilized $(\lambda_{1} < 0)$.
With the further increase of $K$, the largest Floquet multiplier becomes a complex value.
It is precisely at this point that the leading Floquet exponent $\lambda_{1}$ has a minimum value 
(see Fig. \ref{fig:fig5.eps}(b) for the case of $R=0.2$).
Then the pair of complex conjugates move symmetrically and walk out of the unit circle so that
the periodic orbit is destabilized. 
For $K \rightarrow \infty$, the largest Floquet multiplier approaches 
$(\Lambda_{1},\Omega_{1})=(1,0)$ very slowly.
In the same manner with Fig. \ref{fig:fig8.eps}(a), 
the largest Floquet multiplier in Fig. \ref{fig:fig8.eps}(b) (at $R=0.6$)
moves toward the zero point as $K$ increases, until it meets the inner circle of radius $R$.
Then, it jumps to the opposite side of the inner circle and becomes a complex value.
In some finite interval of $K$,
the pair of complex conjugates follow  
a curved line, which is very close to the inner circle.
The loci of the largest Floquet multiplier in this interval explain 
the presence of a flat bottom in Fig. \ref{fig:fig5.eps}(b) for the case of $R=0.6$.
(When the Floquet multiplier locates on the inner circle of radius $R$, 
the corresponding Floquet exponent is $\lambda = \ln |\Lambda+i\Omega|/\tau = \ln (R)/\tau = \lambda^{*}$.) 
Finally, the pair walk out of the unit circle and approach $(\Lambda_{1},\Omega_{1})=(1,0)$.

\begin{figure}[t!]
\includegraphics{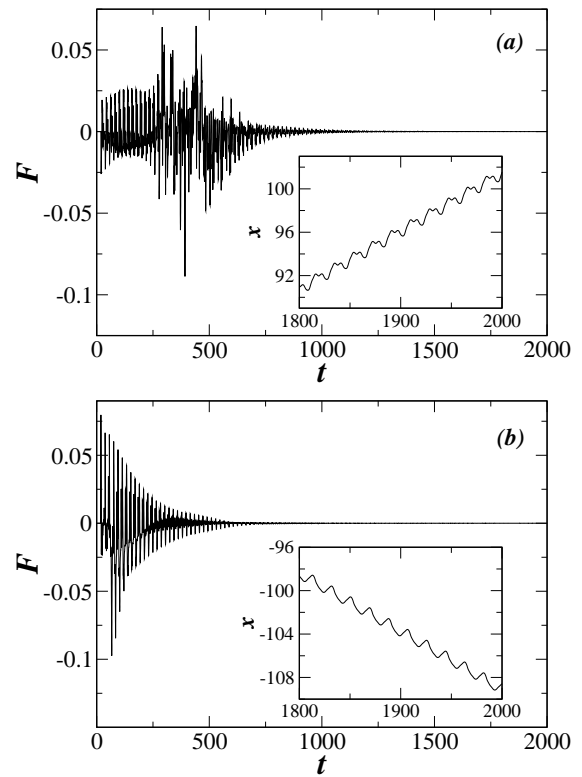}
\caption{Stabilized period-2 orbits (insets) and the dynamics of the control signal $F$ 
at $K=0.35$, $R=0.8$, and $\tau=2T=4\pi/\omega_0$;
(a) the positive current from the initial condition $(x_0,v_0)=(0.0,0.0)$, and
(b) the negative current from $(0.0,0.1)$.}
\label{fig:fig10.eps}
\end{figure}

\section{CONTROL OF TRANSPORT PROPERTIES}

With the results of linear stability analysis of periodic orbits,
we can obtain a desired transport property of the system (i.e., a regular current with a desired mean velocity)
by choosing the control parameters $(K,R,\tau)$ where the corresponding UPO is stabilized.
For simple and efficient application of the EDF method,
it is hoped that each UPO has its own stabilized region of control parameters,
in which the other UPOs still remain in an unstable state.
Some of the UPOs have their own stabilized regions of control parameters.
These orbits are the period-1 confined, the period-2 negative, the period-3 positive, 
and the period-4 negative currents.
For the cases of the period-1 positive, the period-1 negative, and the period-2 positive currents,
the stabilization of each periodic orbit is rather complex
because they do not have their own stabilized regions of control parameters.
The stabilized region of the period-1 positive and the negative currents always overlaps with that of
the period-1 confined current, and the stabilized region of the period-2 positive current overlaps with
that of the period-2 negative current. 

\begin{figure}[t!]
\includegraphics{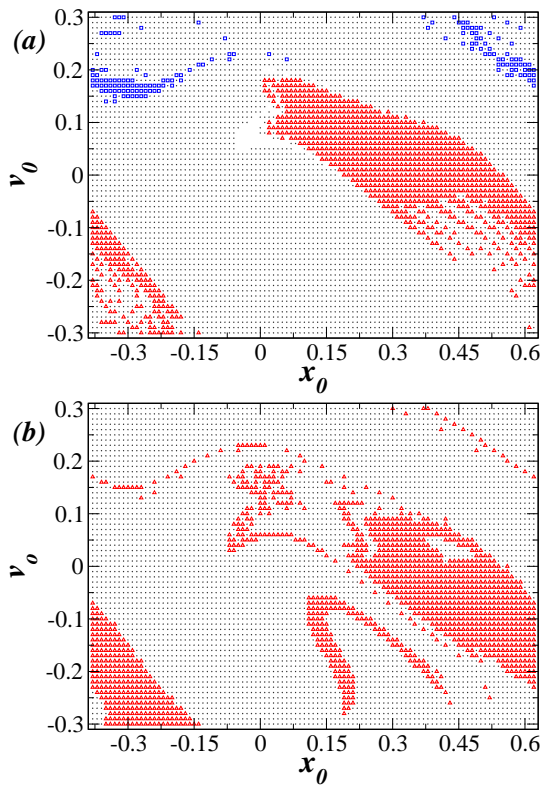}
\caption{(Color online) (a) Basins of period-1 orbits; 
the initial points marked by $\Box$ (blue), $\bigtriangleup$ (red), and $\cdot$ (black) are the basins of the positive, the negative,
and the confined currents, respectively.
(b) Basins of period-2 orbits; 
the initial points marked by $\cdot$ (black) and $\bigtriangleup$ (red) are the basins of the positive and the negative currents, respectively.}
\label{fig:fig11.eps}
\end{figure}

Now, we are interested in the multistable phenomenon that 
more than one UPOs are stabilized at the same control parameters ($K,R,\tau$).
At control parameters, $K=0.67$, $R=0.8$, and $\tau=T=2\pi/\omega_0$, all of period-1 orbits are stabilized.
In Fig. \ref{fig:fig4.eps}, 
each of the leading Floquet exponents of the period-1 orbits at theses parameters
is marked by a black circle and all of them are less than zero.  
Also, all of period-2 orbits are stabilized at $K=0.35$, $R=0.8$, and $\tau=2T=4\pi/\omega_0$. 
Each of the leading Floquet exponents of the period-2 orbits at these parameters is marked by a black circle in Fig. \ref{fig:fig5.eps}.
In the unperturbed system ($K=0$), all of the initial conditions evolve into
a chaotic current in the same manner.
However, the system controlled by the EDF method shows different currents
for different initial conditions ($x_0,v_0$), at control parameters where the multistable phenomenon is observed.

Before considering the numerical integration for obtaining dynamics 
of the system controlled by the EDF method,
we rewrite the control signal $F(t)$ given in Eq. (\ref{eq:5}) into a more convenient form

\begin{eqnarray}\label{eq:13}
 F(t)&=&K\big((1-R)S(t-\tau)-\dot{x}(t)\big), \\
 S(t)&=&\dot{x}(t)+RS(t-\tau) \nonumber,
\end{eqnarray}

\noindent where $S(t)=\sum_{m=0}^{\infty}R^{m}\dot{x}(t-m\tau)$ for an equivalent equation with Eq. (\ref{eq:5}) (see Ref. \cite{Pyragas2}).
In the following numerical integrations, we set $S(t^{'})=0$ for $t^{'}$ in the interval $[-\tau,0]$
and initialize $S(t^{'})=\dot{x}(t^{'})/(1-R)$ for $t^{'}$ in the interval $[0,\tau]$.
Then, the system is not perturbed $(F=0)$ for $t$ in the interval $[0,\tau]$ and perturbed by the control signal from 
$t=\tau$.
In Fig. \ref{fig:fig9.eps}, we have plotted three stabilized period-1 orbits that evolved from different initial conditions
and the dynamics of the control signal $F$ at the same control parameters, $K=0.67$, $R=0.8$, and $\tau=T=2\pi/\omega_{0}$.
For each initial condition, we have integrated Eqs. (\ref{eq:4}) and (\ref{eq:13}).
In Fig. \ref{fig:fig10.eps}, we have shown two stabilized period-2 orbits that evolved from different initial conditions
and the dynamics of the control signal $F$ at $K=0.35$, $R=0.8$, and $\tau=2T=4\pi/\omega_{0}$.
The multistable phenomenon shows that the EDF method can be utilized for separation of particles
in the deterministic inertia ratchet system.
Via the EDF method, we can separate particles in the deterministic inertia ratchet system as their initial conditions vary.
In Fig. \ref{fig:fig11.eps}, we have investigated the basins of period-1 and period-2 orbits.
We have integrated Eqs. (\ref{eq:4}) and (\ref{eq:13}) with the initial condition $(x_0,v_0)$,
where $x_0$ is distributed in one well of the potential, $x_0 \in (-0.38,0.62)$,
and $v_0$ is confined to the ranges of velocities in the unperturbed system, $v_0 \in (-0.3,0.3)$.
As shown in Fig. \ref{fig:fig11.eps}(a), the basins of the period-1 positive, the negative, and the confined currents
are marked by $\Box$ (blue), $\bigtriangleup$ (red), and $\cdot$ (black), respectively.
The basins of the period-2 positive and the negative currents are marked by $\cdot$ (black) and $\bigtriangleup$ (red),
respectively, in Fig. \ref{fig:fig11.eps}(b).

\section{CONCLUSIONS}

We have studied the control of transport properties in the deterministic inertia ratchet system
via the extended delay feedback method.
We have controlled a chaotic current of the unperturbed system to a regular current, 
which has a desired mean velocity.
To obtain the control parameters in which the corresponding unstable periodic orbit is stabilized,
we solve the leading Floquet exponent in the presence of the extended delay feedback.
With the results of leading Floquet exponents as a function of control parameters,
we have obtained a desired regular transport property of the system.
Also, we have observed the multistable phenomenon that 
more than one unstable periodic orbits are stabilized at the same control parameters and
we have shown that the extended delay feedback method can be utilized for separation of particles as a particle's initial condition varies.

\begin{acknowledgments}
The authors thank Dr. S. Rim for valuable discussions.
This study was supported by the Creative Research Initiatives
(Center for Controlling Optical Chaos) of MOST/KOSEF.
\end{acknowledgments}


\begin{thebibliography}{99}

\bibitem{Reimann}P. Reimann, Phys. Rep. \textbf{361}, 57 (2002).

\bibitem{Astumian}R. D. Astumian and P. H\"{a}nggi, Physics Today \textbf{55}, 33 (2002).

\bibitem{Astumian2}R. D. Astumian and M. Bier, Phys. Rev. Lett. \textbf{72},
1766 (1994); R. D. Astumian and M. Bier, Biophys. J. \textbf{70}, 637 (1996);
M. Porto, M. Urbakh, and J. Klafter, Phys. Rev. Lett. \textbf{85}, 491 (2000);
J. V. Hern\'andez, E. R. Kay, and D. A. Leigh, Science \textbf{306}, 1532 (2004).

\bibitem{Zapata}I. Zapata, R. Bartussek, F. Sols, and P. H\"{a}nggi, Phys. Rev.
Lett. \textbf{77}, 2292 (1996);
C. C. de Souza Silva, J. V. de Vondel, M. Morelle, and V. V. Moshchalkov, Nature \textbf{440}, 651 (2006).

\bibitem{Grifoni}P. Reimann, M. Grifoni, and P. H\"{a}nggi,
Phys. Rev. Lett. \textbf{79}, 10 (1997);
M. Grifoni, M. S. Ferreira, J. Peguiron, and J. B. Majer,
{\it ibid}. \textbf{89}, 146801 (2002).

\bibitem{Lee}K. H. Lee, Appl. Phys. Lett. \textbf{83}, 117 (2003);
D. E. Shal\'om and H. Pastoriza, Phys. Rev. Lett. \textbf{94}, 177001 (2005);
M. Beck, E. Goldobin, M. Neuhaus, M. Siegel, R. Kleiner, and D. Kolle, {\it ibid}. \textbf{95}, 090603 (2005);
K. H. Lee, J. Korean Phys. Soc. \textbf{47}, 288 (2005).

\bibitem{Rousselet}J. Rousselet, L. Salome, A. Ajdari, and J. Prost, Nature \textbf{370}, 446 (1994).

\bibitem{Bartussek}R. Bartussek, P. Reimann, and P. H\"{a}nggi,
Phys. Rev. Lett. \textbf{76}, 1166 (1996);
T. E. Dialynas, K. Lindenberg, and G. P. Tsironis,
Phys. Rev. E \textbf{56}, 3976 (1997).

\bibitem{Magnasco}M. O. Magnasco, Phys. Rev. Lett. \textbf{71}, 1477 (1993);
I. Der\'{e}nyi and T. Vicsek, {\it ibid}. \textbf{75}, 374 (1995).

\bibitem{Barbi2}M. Barbi and M. Salerno, Phys. Rev. E \textbf{63}, 066212 (2001).

\bibitem{Savelev}S. Savel'ev, F. Marchesoni, P. H\"{a}nggi, and F. Nori, Phys. Rev. E \textbf{70}, 066109 (2004).

\bibitem{Mateos2}M. Kostur, P. H\"{a}nggi, P. Talkner, and J. L. Mateos, Phys. Rev. E \textbf{72}, 036210 (2005).

\bibitem{Son2}W.-S. Son, Y.-J. Park, J.-W. Ryu, D.-U. Hwang, and C.-M. Kim, to be published in J. Korean Phys. Soc.


\bibitem{Ott}E. Ott, C. Grebogi, and J. A. Yorke, Phys. Rev. Lett. \textbf{64}, 1196 (1990).

\bibitem{Shinbrot}T. Shinbrot, C. Grebogy, E. Ott, and J. A. Yorke, Nature \textbf{363}, 411 (1993);
{\it Handbook of Chaos Control}, edited by H. G. Schuster (Willey-VCH, Weiheim, 1999);
S. Boccaleti, C. Grebogi, Y.-C. Lai, H. Mancini, and D. Maza, Phys. Rep. \textbf{329}, 103 (2000).

\bibitem{Pyragas}K. Pyragas, Phys. Lett. A \textbf{170}, 421 (1992).

\bibitem{Ushio}T. Ushio, IEEE Trans. Circuits Syst. I, Fundam. Theory Appl. \textbf{43}, 815 (1996);
W. Just, T. Bernard, M. Ostheimer, E. Reibold, and H. Benner, Phys. Rev. Lett. \textbf{78}, 203 (1997);
H. Nakajima, Phys. Lett. A \textbf{232}, 207 (1997);
W. Just, E. Reibold, H. Benner, K. Kacperski, P. Fronczak, and J. A. Ho{\l}yst, Phys. Lett. A \textbf{254}, 158 (1999).

\bibitem{Socolar}J. E. S. Socolar, D. W. Sukow, and D. J. Gauthier, Phys. Rev. E \textbf{50}, 3245 (1994).

\bibitem{Kittel}A. Kittel, J. Parisi, and K. Pyragas, Phys. Lett. A \textbf{198}, 433 (1995);
H. G. Schuster and M. B. Stemmler, Phys. Rev. E \textbf{56}, 6410 (1997);
K. Pyragas, Phys. Rev. Lett. \textbf{86}, 2265 (2001).

\bibitem{Bleich}M. E. Bleich and J. E. S. Socolar, Phys. Lett. A \textbf{210}, 87 (1996);
W. Just, E. Reibold, K. Kacperski, P. Fronczak, J. A. Ho{\l}yst, and H. Benner, Phys. Rev. E \textbf{61}, 5045 (2000);
K. Pyragas, Phys. Rev. E \textbf{66}, 026207 (2002).

\bibitem{Mateos}J. L. Mateos, Phys. Rev. Lett. \textbf{84}, 258 (2000).

\bibitem{Son}W.-S. Son, I. Kim, Y.-J. Park, and C.-M. Kim, Phys. Rev. E \textbf{68}, 067201 (2003).

\bibitem{Barbi}M. Barbi and M. Salerno, Phys. Rev. E \textbf{62}, 1988 (2000).

\bibitem{Pyragas2}K. Pyragas, Phys. Lett. A \textbf{206}, 323 (1995).

\end{thebibliography}
\end{document}